# Covid-19 lockdown improves air quality in Morocco

Imane SEKMOUDI [a]*, Kenza KHOMSI[b], Soufiane FAIEQ[c, d], Laila IDRISSI [a],

[a] Process and Environment Engineering Laboratory, Faculty of Sciences and Technologies, Mohammedia Hassan II University, Casablanca, Morocco

[b] Direction of National Meteorology, Casablanca, Morocco

[c] Univ.Grenoble Alpes , CNRS,Grenoble INP( Institute of Engineering Univ.Grenoble Alpes) , LIG, 3800 Grenoble, France

[d] LRIT Associated Unit to CNRST (URAC 29), Faculty of Sciences, Mohammed V University in Rabat, Rabat, Morocco

**Keywords**: Covid-19, lockdown, Morocco, Satellite data, $PM_{2,5}$, $NO_2$.

**Abstract**

Due to the Covid-19 pandemic, almost all non-essential activities in Morocco have been halted since March 15, 2020. In this sense, several measures have been taken to mitigate the effects of this pandemic. Morocco has been on lockdown since March 20, 2020. The main objective of this paper is to study the effects of the lockdown measures on air quality, by presenting the spatio-temporal evolution of $PM_{2,5}$ and $NO_2$ pollutants, based on satellite data from Sentinel 5P and MERRA. Then, these results are compared to the data predicted by the Copernicus Atmosphere Monitoring Service (CAMS). Results show that, on one hand, Morocco has experienced an improvement in air quality by a significant reduction in $NO_2$ and a relative reduction in $PM_{2,5}$. On the other hand, we demonstrate that the particulate pollution in Morocco is partly due to synoptic weather patterns, and that a significant part of $NO_2$ emissions is incoming mainly from the outside northern borders of Morocco. Finally we show that $PM_{2.5}$ CAMS global forecasts underestimate the observed MERRA data.

*Corresponding author at: Process and Environment Engineering Laboratory, Faculty of Sciences and Technologies, Mohammedia Hassan II University, Casablanca, Morocco

E-mail address: imane.sekmoudi@gmail.com (I. SEKMOUDI).

1. **Introduction**

The new SARS-CoV-2 coronavirus was first discovered in China at its epicenter Wuhan. On March 02, 2020, Morocco announced the first confirmed case of Covid-$_{19}$. Since that day, to limit the spread of this pandemic, Moroccan authorities have taken several measures and proactive decisions. On March 11, 2020, Morocco created an Economic Vigilance Committee and then informed the public by all possible media about the preventive measures to limit the spread of the virus. The declaration of "state of health emergency" was announced for the first time on March 20, 2020, for a period of one month, and then it was extended three times until July 10, 2020.

The main objective of the present paper is to study the effects of Covid-$_{19}$ induced lockdown on air quality by presenting the spatio-temporal evolution of $PM_{2.5}$ and $NO_2$ pollutants as indicators of air quality, using data from MERRA-2 and the sentinel 5P satellites. Then, these results were compared to data predicted by the CAMS global forecasts data.

This paper is structured as follows. Section 2 present the international and national experiences of covid-19 lockdown and air pollution, section 3 describes the data used, the period, and the study area. In section 4, we present the results of the spatio-temporal evolution of $PM_{2.5}$ and $NO_2$ concentrations as well as the comparison of MERRA $PM_{2.5}$ data and that predicted by CAMS global forecasts. In section 5, we present the conclusion and the recommendations of the authors.

## 2. Covid-19 pandemic and air pollution

Since the declaration of the Covid-19 epidemic as a Public health emergency of international concern by the World Health Organization (WHO) (Sohrabi et al., 2020) several countries have implemented flight restrictions and preventive measures(Anderson et al., 2020). Morocco took the initiative to control citizens from countries affected by Covid-19 and then suspended the majority of industrial and commercial activities. In addition, the transportation and movement of citizens were limited only to necessary needs of work, medical consultations and purchases of food needs, or other compulsory necessities. In the same way, all social activities that result in mass gathering such as ceremonies, educational establishments, religious festivals, sports competitions were prohibited (P, 2020).

At the international level, thanks to the reduction in activities (traffic and industrial activities) during the Covid-19 lockdown, several studies have reported a reduction in air pollution. Starting with China, in Wuhan (Isaifan, 2020) and in 44 cities in northern China, both studies recorded an improvement in the air quality index (Baldasano, 2020). Similar results have been reported in Italy, France, Spain, Germany, Bangladeshl and the USA (LARNAUD, 2020; Adoo-kissi-debrah, 2020; Baldasano, 2020;Ogen, 2020 ;Roy et al., 2020; Berman and Ebisu, 2020). This was also observed in the most polluted cities in the world, notably Bangalore, Beijing , Bangkok, Delhi and Nanjing, as well as in the main shopping centers of the world, notably New York, London, Paris, Seoul, Sydney and Tokyo (Sharma et al., 2020).

Another study compared the state of air quality before the crisis with the current situation showing that blocking forced and industrial activities of Covid-19 may have saved more lives by preventing ambient air pollution than by preventing infection. On the one hand, in 2016, at the global level, the death rate caused by ambient air pollution contributed to 7.6% of all deaths, on the other hand the death rate from this infection did not exceed 3, 4% worldwide (Isaifan, 2020).

At the national level, a study carried out during this pandemic estimated that the concentrations of $PM_{10}$, $SO_2$ and $NO_2$ will continue to decrease and remain at a minimum level during the lockdown period and they concluded that the reduction in the concentrations of these pollutants can be mainly attributed to drastic measures limiting human movement and industrial activities (Otmani et al., 2020).

On the other hand, In China, a study has shown that serious episodes of air pollution are not avoided by reducing activities (traffic and industrial activities) during Covid-$_{19}$, especially when the weather is unfavorable (Wang et al., 2020). More we must know the sustainability of the benefits of reducing air pollution thanks to Covid-$_{19}$ by looking at the situation after the crisis of this pandemic (Isaifan, 2020).

3. **Data and methods**

   3.1. **Sentinel 5P**

Sentinel 5p is the first Copernicus mission satellite devoted to monitoring the atmosphere, it was launched on October 13, 2017 (Shikwambana et al., 2020). For this paper, the TROPOMI Offline of daily $NO_2$ concentration in the processing level 2 datasets were used for the period from January 22, 2020 to April 22, 2020 for the Moroccan area and its surroundings.

Several studies have used the sentinel 5P source for the observation of $NO_2$ during the period of the Covid-19 pandemic, for example in South Africa (Shikwambana et al., 2020) and in 66 administrative regions of Italy, Spain, France and Germany (Ogen, 2020).

### 3.2. Modern-era retrospective analysis for research and applications version 2 (MERRA-2)

MERRA-2 provides atmospheric data dating back from 1980 to date. It is the first long-term global reanalysis to assimilate space-based observations of aerosols and represent their interactions with other physical processes in the climate system (Pawson, 2019). The MERRA-2 is a National Aeronautics and Space Administration (NASA) atmospheric reanalysis for the satellite era using the Goddard Earth Observing System Model, Version 5 (GEOS-5) with its Atmospheric Data Assimilation System (ADAS), version 5.12.4 (Gelaro et al., 2017).

The MERRA-2 data used in this paper relates to particulate matter less than 2,5 μm from February 20, 2020 to April 20, 2020 in Morocco with a Spatial resolution of 0.5 x 0.625 °.

The choice of monitoring $PM_{2.5}$ concentrations in this study is based on its environmental and especially health impact, and on the study which indicates that long-term exposure to this pollutant may be one of the most important contributors to the deaths exacerbated by the Covid-19 virus and possibly worldwide (Xiao Wu et al., 2020).

### 3.3. Copernicus Atmosphere Monitoring Service (CAMS) global forecasts

CAMS which is implemented by the European Centre for Medium-Range Weather Forecasts (ECMWF), is a component of the European Earth Observation programme Copernicus. CAMS global near-real time (NRT) service provides daily analyses and forecasts of reactive trace gases, greenhouse gases and aerosol concentrations (ECMWF COPERNICUS, 2012). The predicted data used in this study concern the $PM_{2.5}$ concentration at the surface level.

### 3.4. ERA5

ERA5 provides hourly estimates of a large number of atmospheric, land and oceanic climate variables. It includes information about uncertainties for all variables at reduced spatial and temporal resolutions. The data cover the Earth on a 30km grid and resolve the atmosphere using 137 levels from the surface up to a height of 80km (ECMWF COPERNICUS, n.d.).

ERA5 data was used to acquire the mean sea level pressure on March 20, 2020.

## 4. Results and discussion

### 4.1. Spatio-temporal evolution of $PM_{2,5}$ and $NO_2$

In this part, we have chosen the presentation of the spatio-temporal evolution of two pollutants indicative of air quality which are $PM_{2,5}$ and $NO_2$. This presentation runs from a period before the appearance of the first confirmed case of Covid-19 in Morocco (from January for PM2.5 and February for NO2) until April 2020.

### 4.1.1. PM$_{2,5}$

Regarding PM$_{2.5}$, the data used come from the MERRA-2 project and the duration of the study extends from February 20 to April 20, 2020. Figure 1 clearly shows that we have a flow of particles from the south and south-eastern of Morocco.

In general, the maps show a relative decrease in PM$_{2.5}$ concentrations thanks to the reduction of industrial activities and transport during the lockdown period in Morocco.

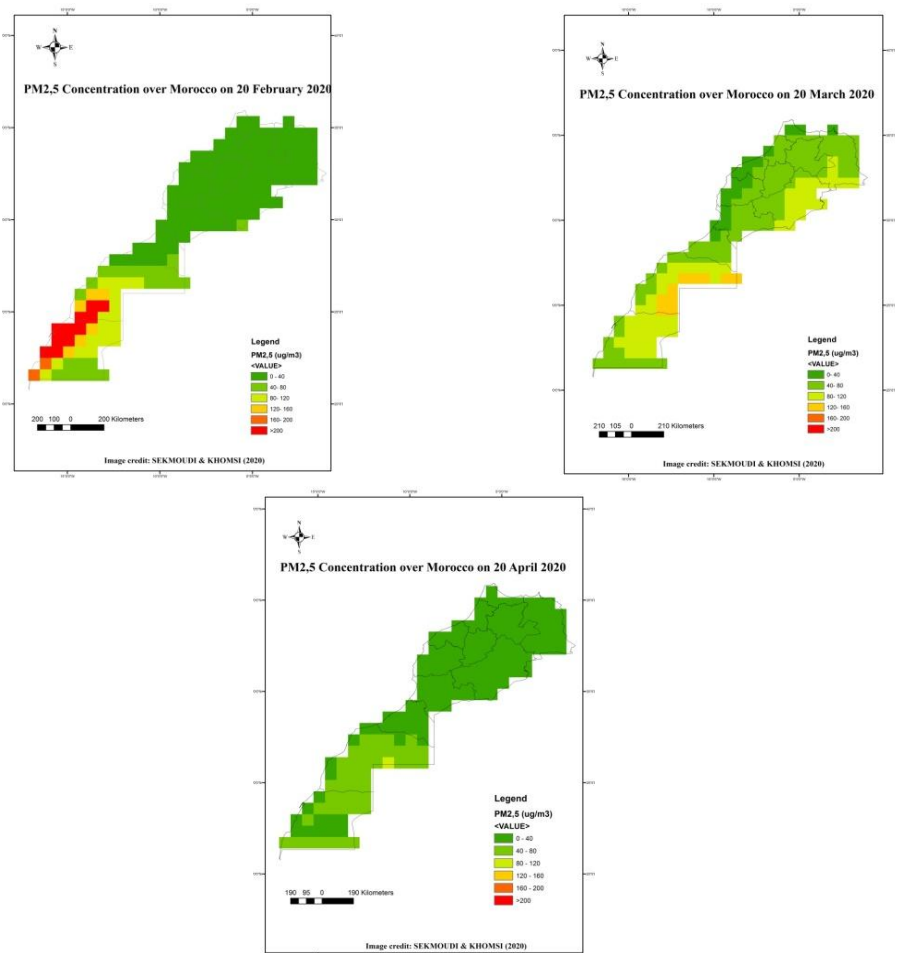

Figure 1: Spatio-temporal evolution of PM$_{2,5}$ Concentration over Morocco

### 4.1.2. NO$_2$

For nitrogen dioxide, we used Sentinel 5P data. Figure 2 illustrates the spatio-temporal evolution of the NO$_2$ concentration from January 22 to April 22, 2020. These maps show a significant decrease in NO$_2$ concentrations and give a vision of the contribution of emissions from the external borders of northern Morocco, Which confirms the study highlighting the predominant influence of transboundary pollution on the air quality of the northern cities of Morocco (Benchrif et al., 2018).

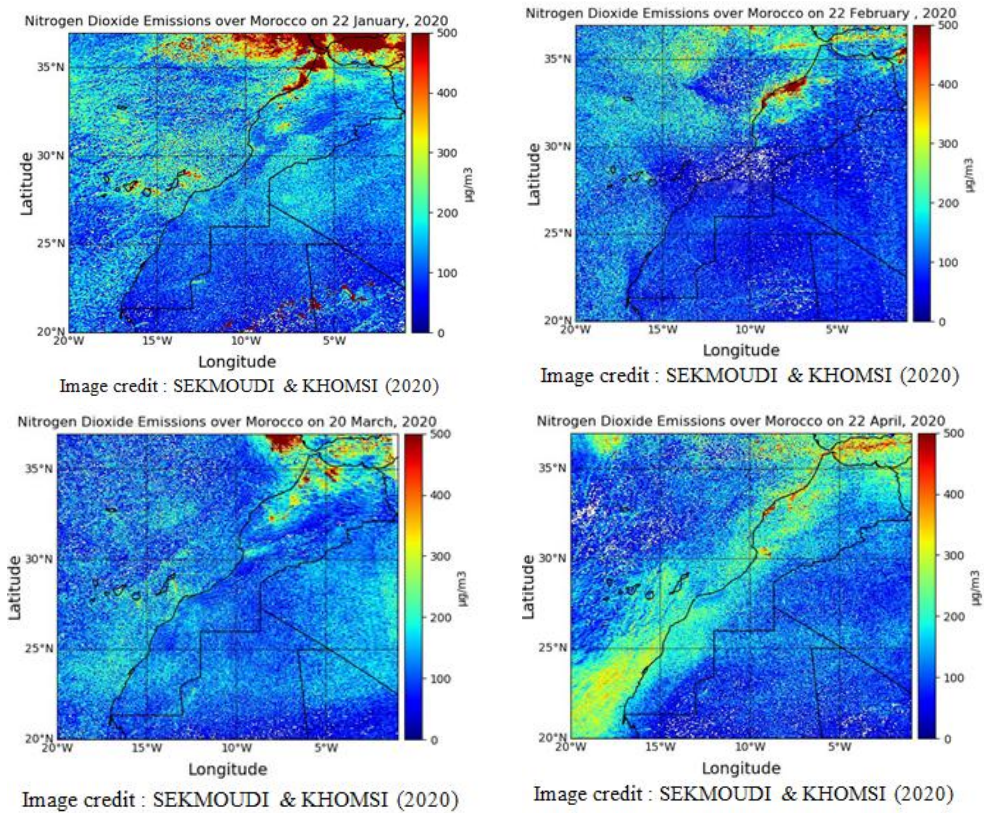

Figure 2: Spatio-temporal evolution of NO$_2$ Concentration over Morocco

### 4.2. Comparison of observed and predicted data

Figure 3 shows the concentrations of PM$_{2.5}$ forecasts data from CAMS during the lockdown period. The difference between the map of February 20, 2020 and that of April 20, 2020 shows a decrease in the concentration of PM$_{2.5}$. However the map of March 20, 2020 shows a situation of stagnation of dust because of the high pressure in the north of morocco which is illustrated in figure 4.

We conclude from the comparison of data predicted by CAMS (Figure 3) and that observed by MERRA (Figure 1), that the data predicted underestimates the observed MERRA data.

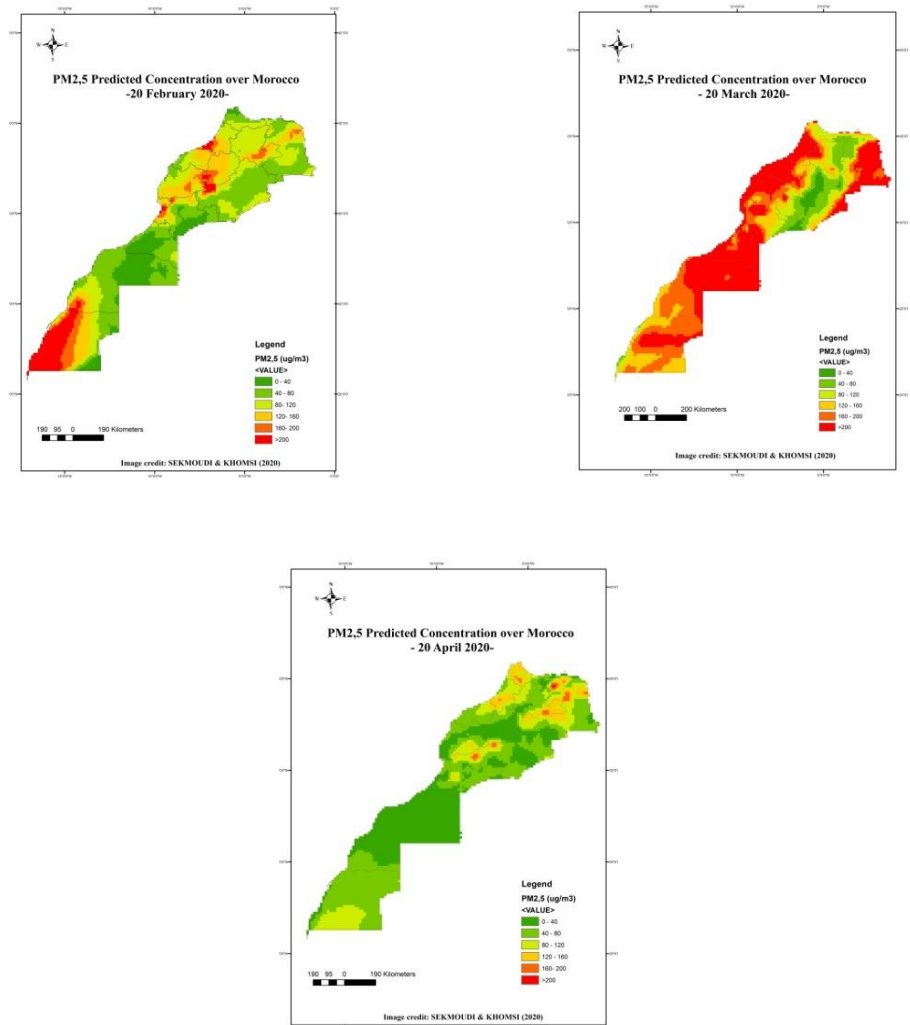

Figure 3: Spatio-temporal evolution of PM$_{2,5}$ predicted concentration over Morocco

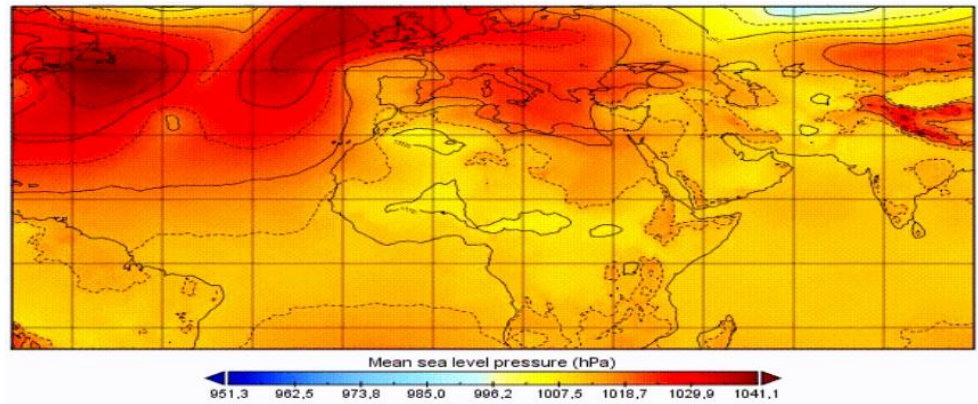

Figure 4: Mean sea level pressure on 20 March 2020

### 4.3. Improvement in air quality during the Covid-19 lockdown

The results of this study showed a significant reduction in air pollution, which remains relative to the state of each city. This result goes in the same direction with the study concluded that the reduction of air pollution in Morocco is relative for each city according to the sources of emission, and therefore it is necessary to study each city in isolation in order to improve air quality (Croitoru and Sarraf, 2017).

On the other hand, this improvement in air quality remains to be seen if the lockout is extended. However, these reductions are expected to be temporary as levels should rise once the situation returns to normal(Sharma et al., 2020).

### 5. Conclusion

In conclusion, in Morocco, the Covid-19 lockdown has a positive effect on air quality. The reduction in $PM_{2.5}$ and $NO_2$ concentrations during the lockdown period can be mainly attributed to measures taken by the Moroccan state by limiting human activities, namely transport and industrial activities. This could be a lesson in order to strengthen national and international environmental strategies and adopt new attitudes.

The authors show that the particulate pollution in Morocco is partly due to the synoptic weather factors, and that a significant part of the $NO_2$ emissions is mainly incoming from the northern borders of Morocco. The authors also showed that the $PM_{2.5}$ concentration predicted by CAMS underestimates that observed by MERRA.

In order to remedy this situation and improve air quality, the authors recommend on the one hand, the establishment of a remote work integration plan in the private sector and the state to minimize transportation, the encouragement of the use of electronic vehicles, and the enhancement and encouragement of the use of adequate public transport. On the other hand, the encouragement of companies to convert their work remotely as possible, also the increase of the number of event days (day without car, bike day, ...) must be taken into account as well as more funds should be provided to scientific research, specifically that related to air quality.

**Declaration of competing interest**

The authors declare no conflict of interest.

**CRediT authorship contribution statement**

Imane SEKMOUDI: Conceptualization, Writing - original draft, Data analysis. Kenza KHOMSI: Conceptualization, Supervision. Soufiane FAIEQ: review & editing, Supervision. Laila IDRISSI: Supervision.

**Acknowledgements**

We acknowledge the Analysis Infrastructure (Giovanni) and ESA for providing us the MERRA-2 data, the Sentinel-5 P/TROPOMI and ERA5 products. The authors gratefully acknowledge the Copernicus programme.

**Funding**

This research did not receive any specific grant from funding agencies in the public, commercial, or not-for-profit sectors.

**Ethical approval**

Not required.